\documentclass[a4paper]{jpconf}
\usepackage{graphicx}
\usepackage{lineno}
\begin{document}
\title{Recent results from the Pierre Auger Observatory}

\author{Esteban Roulet, for the Pierre Auger Collaboration}

\address{CONICET, Centro At\'omico Bariloche, Bustillo 9500, Bariloche, 
8400, Argentina}

\ead{roulet@cab.cnea.gov.ar}
\begin{abstract}
The main results from the Auger Observatory are described. A
steepening of the spectrum is observed at the highest energies, supporting the
expectation that above $4\times 10^{19}$~eV 
the cosmic ray energies are significantly degraded by 
interactions with the CMB photons
(the GZK effect). This is further supported by the correlations observed 
 above $6\times 10^{19}$~eV with the
distribution of nearby active galaxies, which also show the
potential of Auger to start the era of charged particle astronomy. The lack of
observation of photons or neutrinos strongly disfavors top-down
models, and these searches may approach in the long term  the sensitivity
required to test the fluxes expected from the secondaries of the very 
same GZK process.  Bounds on the anisotropies at EeV energies
contradict hints from previous  experiments that suggested a large excess 
from regions near the Galactic centre or  the presence of
  a dipolar type modulation of the cosmic ray flux.
\end{abstract}

{\bf Introduction: }
After having been studied for almost a century, cosmic rays (CRs) still offer
many puzzling results, particularly at the highest energies which are now
being explored at the Pierre Auger Observatory.
From $10^9$~eV up to about $10^{20}$~eV the spectrum of cosmic particles is
essentially a power law  d$N$/d$E\sim
E^{-\alpha}$, with $\alpha \simeq 3$, with some small but relevant
breaks at the so-called knee (at about $5\times 10^{15}$~eV), the second knee
(slightly above $10^{17}$~eV) and the ankle (at about $3\times 10^{18}$~eV).
The steeply falling nature of the CR spectrum implies that  
although the fluxes are large at low energies, i.e. of order
1 particle per cm$^2$ per second in the GeV range, they become extremely 
small in the
highest energy end, e.g. of about 1~particle per km$^2$ per century above $6\times
10^{19}$~eV. 
This implies that huge detectors are required to gather sizeable statistics in
this last regime. 
 
The non-thermal nature of CRs is what led Fermi to speculate that their
acceleration was the result  of stochastic processes involving charged
particles in the presence of astrophysical magnetic fields, what later evolved
into the scenario of diffusive acceleration in shock waves. In particular, the
bulk of the CRs of low energies may be produced in this way in galactic
supernova explosions. But the highest energies are so extreme that only a few
sites are large enough and have sufficiently strong magnetic fields  to be
(marginally) able to produce them, with possible candidates being active
galactic nuclei and gamma-ray bursts.
If indeed the highest energy cosmic rays are extragalactic, as is also
suggested by the lack of observed anisotropies associated with the galactic
plane, a further issue is that inelastic collisions with CMB photons are
expected to degrade the CR energies as they propagate. If CRs are protons this
takes place by photopion production, while if they are heavier nuclei the
dominant process is photodisintegration. In both cases losses become
important above a certain energy threshold, which coincidentally is $\sim
5\times 10^{19}$~eV for both protons and Fe nuclei, and is somewhat smaller
for lighter nuclei. As a consequence of this, the spectrum is expected to
steepen above this so-called GZK threshold even if the sources  continue to be
powerful. Until very recently the existence, or otherwise, 
 of this suppression was strongly
debated, with results from the AGASA instrument (detecting air showers at
ground level with scintillators) not observing any signs of it, while the HiRes
experiment (looking to the fluorescence emitted in the air by nitrogen
molecules excited by the passage of the shower) found indications in favor
of it. 

This situation led in the past to the proposal of many exotic
scenarios to overcome the limitations of the bottom-up acceleration in 
known astrophysical sources, in which CRs were 
produced instead in a  top-down way, e.g. from decays of topological
defects or superheavy relics from the early universe, from `Z-burst' models,
etc.  One of the main distinguishing features of the top-down scenarios is
that a large fraction of photons and neutrinos should be present at the
highest energies. Also exotic physics, such as Lorentz invariance violation,
was sometimes invoked to avoid the prediction of a GZK suppression.
Anyway, the AGASA and HiRes results were based on small number of events
 and, using
different techniques, were  affected by different systematics. To overcome
these difficulties the Auger Observatory was conceived as
a hybrid detector combining the two detection methods
 and covering an area 30 times bigger than that of AGASA.
As is discussed below,  even during the Auger
construction phase which is now almost finished, it proved possible to
obtain  measurements of the  CR spectrum,
composition and arrival directions which already contributed 
 significantly to the  progress in this field.

{\bf The Observatory and recent results:}
The Auger Observatory is located near the town of Malarg\"ue, Argentina, at
 35.2$^\circ$~S latitude and 1400~m a.s.l. 
It is an international 
 collaboration of about 400 scientists from 17 countries.
The Observatory has a hybrid design, with 
the surface detector (SD)  consisting of
 a grid of 1600 water Cherenkov stations with 1.5~km spacing, covering a total
 of 3000~km$^2$, and the fluorescence detector (FD) consisting of 4 buildings
 on the outskirts of the SD array, each one with 6 telescopes 
 that overlook the array covering $30^\circ$ in elevation and  180$^\circ$
 in azimuth. The surface detectors are tanks with 12~tonnes of pure water in
 which Cherenkov light is produced by both the electromagnetic and the muonic
 component of the air showers. This light, after reflection by the
 diffuse liner, is  detected by three 9'' phototubes. Three or
 more nearby triggered stations are required to detect a shower, and while the
 relative timing (with a 25~ns sampling) gives the information on
 the arrival direction, the size of the signal contains  information on the
 shower energy $E$. A fit is indeed performed to timing and signal data to
 reconstruct the location of the shower core and the lateral distribution  of
 the shower. From this the expected signal size at 1000~m from the core,
 $S(1000)$,  is
 obtained, and this is used as an estimator of the shower energy.
Actually, for a given CR energy the showers with different zenith angles 
$\theta$ are being
sampled at ground level in different stages of their development, so that the
relationship between $S(1000)$ and $E$ is $\theta$ dependent. A direct way
to obtain this attenuation effect without relying on shower simulations is to
use the fact that the CR flux is essentially isotropic, so that  above a given
energy  one expects a uniform number of events in bins of equal exposure
(i.e. in bins of cos$^2\theta$ once the detector is fully efficient, which is
the case for $E>3\times 10^{18}$~eV). 
 Hence, looking in each bin of cos$^2\theta$ 
for the value of the 
signal above which there is a certain fixed given number of
events, the $\theta$ dependence of the 
relation between $S(1000)$ and energy is obtained. This 
 allows a quantity  $S_{38}$ to be assigned 
for each value of $S(1000)$ and $\theta$, which is 
the  signal  that would have been observed had the shower arrived
with $\theta=38^\circ$ (the median zenith for the showers with
$\theta<60^\circ$, which is the range in which most analyses rely). 
To calibrate the energies one can use the showers measured with both
SD and FD during clear moonless nights. For these hybrid showers
the energy can be measured almost calorimetrically with FD by fitting
the longitudinal development of the shower, integrating it to include
the tails and accounting also for the unobserved particles in the
shower (neutrinos and muons). However, since what is 
observed by the FD telescopes is
the light emitted by the air molecules, not directly the energy deposited by
the shower particles, the inferred energy depends sensitively on the value
adopted for the
fluorescence yield, which is unfortunately still subject to significant
systematic errors (for instance Auger and HiRes use determinations that differ
by about $\sim 10$\%). It is expected that
 the determination of the
fluorescence yields will improve
soon, allowing   a major source of systematic uncertainty in the
determination of the shower energies to be eliminated.
With this set of hybrid events a relation of the form $E_{FD}=A\; S_{38}^B$
can be fitted to the data, as is shown in fig.~1, and this can
then be used to obtain the energies of all the SD events
 (only about 15\% of the showers are
hybrid). From the inset in the figure it is seen that a 19\% dispersion
remains in this relation, associated mainly to the reconstruction procedure
and to shower-to-shower fluctuations. The systematic
uncertainty in the energy assignment is 22\%.

\begin{figure}[h]
\begin{minipage}{18pc}
\includegraphics[width=15pc]{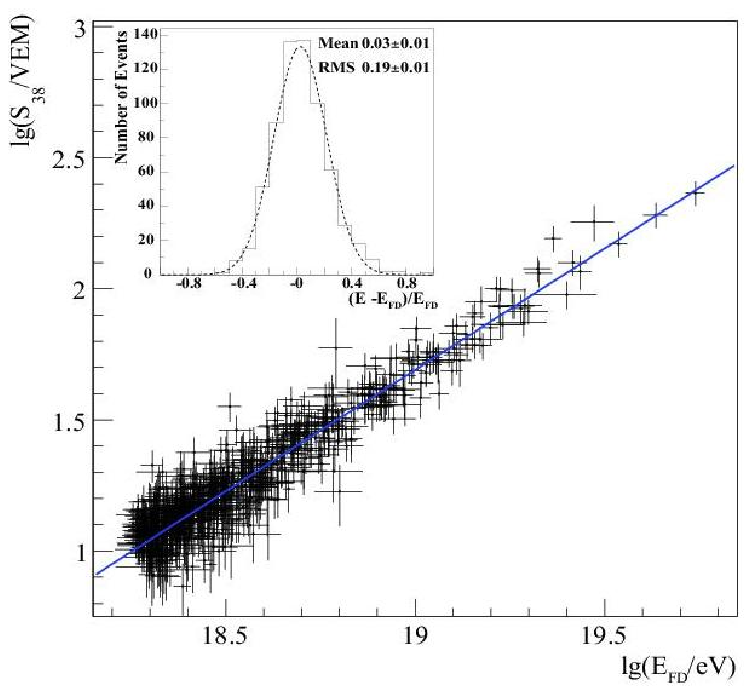}
\caption{\label{cal3}Correlation between the SD energy estimator $S_{38}$ and
  the FD energy for good quality hybrid events. The inset shows the 
  dispersion in $E_{SD}/E_{FD}$.}
\end{minipage}\hspace{1pc}%
\begin{minipage}{19pc}
\includegraphics[width=14pc]{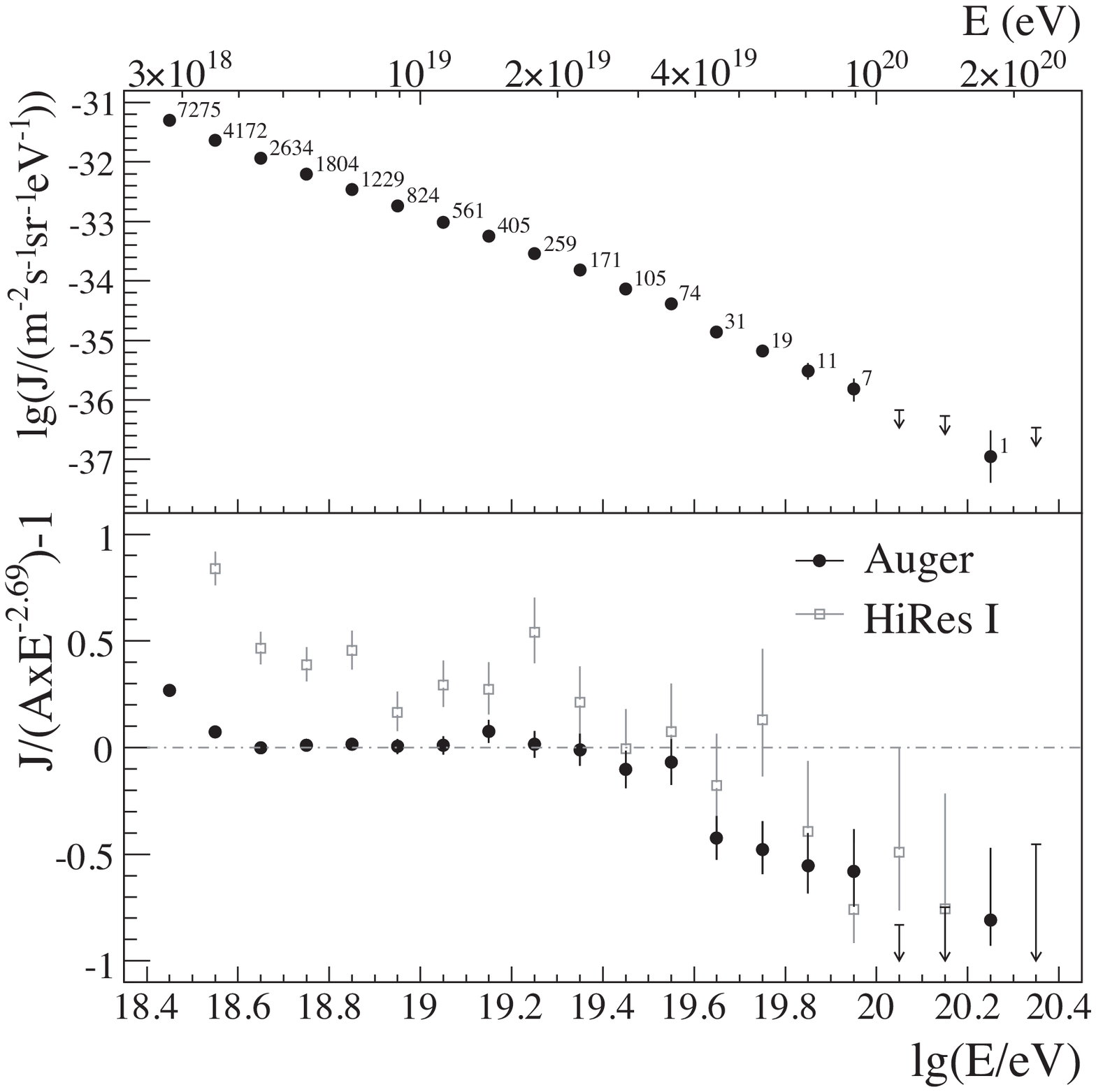}
\caption{\label{spec}Auger differential energy spectrum with statistical 
error bars. Lower panel shows fractional differences of Hires I and Auger data
with respect to a power law with index $2.69$.}
\end{minipage} 
\end{figure}

The energy spectrum obtained \cite{spectrum} 
from the events collected up to August 2007
(about 20000 events above $2.5\times 10^{18}$~eV, with about four times
the exposure of AGASA and twice that of HiRes) is shown in 
fig.~2. To better appreciate the suppression at the highest energies it
 is also shown in the lower panel normalized
 to the power law $E^{-2.69}$ which fits the
spectrum below $4\times 10^{19}$~eV. 
The numbers of events expected if this power law were to hold above $4\times
10^{19}$~eV or  $10^{20}$~eV, would be
$167\pm 3$ and $35\pm 1$, while 69 events and 1 event are observed, clearly
showing the strong suppression present and rejecting the simple power law
extrapolation at more than 6$\sigma$ level.  For comparison also the latest
HiRes  \cite{hires}
results are shown, with the two experiments being in agreement, except for a
possible systematic energy mismatch that is necessary to account for
 the different flux normalizations obtained.

If the suppression in the CR spectrum is indeed due to the interactions with
the CMB photons (and not just due to an unfortunate coincidence with the
maximum attainable energies at the sources) a
 further crucial point is that when looking at energies above the GZK
threshold the only CRs that can reach us are those produced in relatively
nearby  sources (for instance, above $6\times 10^{19}$~eV 90\% of the
protons should come
from less than $\sim 200$~Mpc, while 50\% should come from less than 90~Mpc). 
Hence,  the arrival directions of the highest energy CRs are
 expected  to be correlated with the nearby matter distribution, which is quite
 inhomogeneous.
 Observing this kind of correlations can  be a first step towards 
 doing actual CR astronomy. Note that a major difficulty for charged particle
 astronomy is the fact that deflections in the  galactic
and extragalactic magnetic fields may well be large,  of order $10^\circ
Z(10^{19}\ {\rm eV}/E)$, with $Z$ the CR charge. On the other hand,  
the angular resolution of Auger above $10^{19}$~eV is 
better than one degree.

\begin{figure}[h]
\includegraphics[width=23pc]{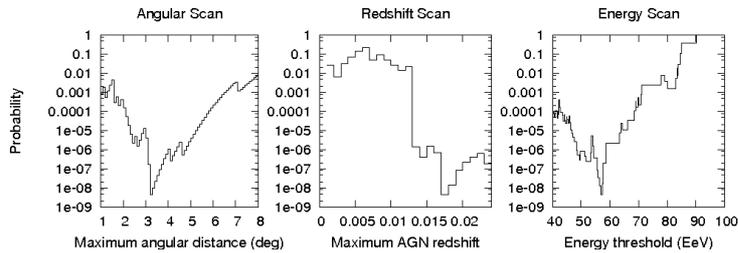}\hspace{0pc}%
\begin{minipage}[b]{14pc}\caption{\label{pplots}Probability for the null
    hypothesis (isotropy) vs. $\psi$, $z_{max}$ and $E_{th}$. In each case the other two parameters are
kept fixed 
at the values leading to the minimum probability ( $\psi =
3.2^\circ$, $ z_{max} =0.017$, $E_{th} = 57$~EeV).}
\end{minipage}
\end{figure}

To test for possible correlations with  extragalactic 
sources the Auger collaboration analyzed the arrival directions of the events 
above $4\times 10^{19}$~eV to look for coincidences with the positions of the
known  nearby (less than 100~Mpc) active galactic nuclei \cite{science,
  longagn}.  Given the unknown
magnetic deflections, the possible systematic uncertainties in energy as 
well as the unknown
CR composition (what would also affect the GZK horizon distance), a scan
over the angle $\psi$ between the events and the AGNs, 
 the maximum AGN redshift considered $z_{max}$  and the threshold energy
$E_{th}$ is performed to search 
for the most significant correlation. The results of this scan are shown in
fig.~3, showing a deep minimum in the probability $P$ of observing a similar
or larger number of correlations arising from isotropic simulated data. This
minimum is obtained for $\psi=3.2^\circ$, $z_{max}=0.017$ (or maximum AGN
distance of $71$~Mpc) and
$E_{th}=57$~EeV (corresponding to
the 27 highest energy events), where 1~${\rm EeV}\equiv 10^{18}$~eV. Only $\sim
10^{-5}$ of the isotropic simulations have a deeper minimum under a 
similar scan.
In particular, for these 27 events 20 are at less than $3.2^\circ$ from an 
AGN closer than
71~Mpc, while only 6 were expected to be found by chance from an isotropic
distribution of arrival directions. A correlation was first
observed in the data obtained before the end of May 2006, with a very similar
set of parameters, 
and fixing that set of parameters  a priori the subsequent 
data up to
August 2007 were studied, confirming the original 
correlation with more than 99\% CL significance in the additional data set
alone.

The map of the arrival directions and of the AGN positions is shown in
fig.~4. A remarkable alignment of several events with the supergalactic plane
(dashed line) is observed, and it is also worth noting that 
two events fall within 3.2$^\circ$ from Centaurus~A, the closest active galaxy.
A further interesting fact is that the energy maximizing the correlation with
AGNs coincides with that maximizing the autocorrelation of the events
themselves \cite{autocor} and is also that for which the spectrum falls to
half of the power law extrapolation from smaller energies (fig.~2).

\begin{figure}[h]
\includegraphics[width=22pc]{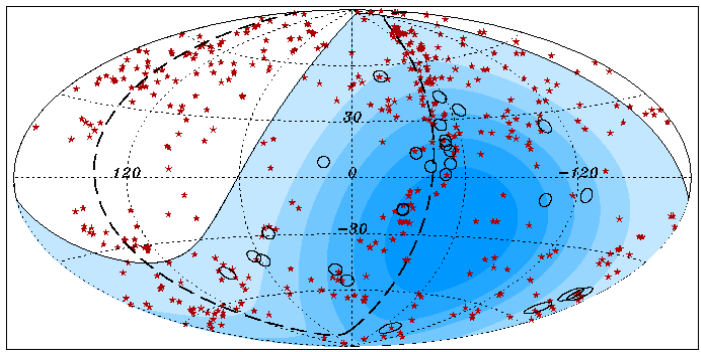}\hspace{3pc}%
\begin{minipage}[b]{10pc}\caption{\label{skymap}Map in galactic coordinates
    with the positions of the AGNs within 71~Mpc (stars) and the 27 events
    with $E>57$~EeV (circles of 3.2$^\circ$ radius). Shading indicates regions
    of equal exposure.}
\end{minipage}
\end{figure}

Let us also mention that a search for correlations with BL Lacs gave negative
results \cite{bllac}.

Another important search performed by Auger was to look for the presence of
photons in the highest energy CRs. One possible signature of photon showers is
their comparatively larger values of $X_{max}$, the column depth of air at
which the longitudinal development reaches a maximum. This is due to  the
slower development of purely electromagnetic showers with respect to hadronic
ones. A first limit was set using this technique in \cite{fdgamma} using
the FD measurements. A more sensitive search for photons was done 
using the full statistical power of the SD
detector and looking at the risetime of the SD signals (which are slower in
muon poor showers) and the 
curvature radius of the front of the showers (which is smaller
in late developing showers). This led to the bounds \cite{gammabound}
displayed in fig.~5, which show in particular 
that above $10^{19}$~eV only less than 2\% of the CRs may be
photons. This excludes most of the top-down model predictions (also shown in
the figure).  The amount of photons expected  (dashed region)
from the decays of neutral pions
produced in the GZK process, if CRs at the highest energies are proton
dominated, is somewhat below present sensitivities, but may be within the
reach of future searches.

\begin{figure}[h]
\begin{minipage}{18pc}
\includegraphics[width=15pc]{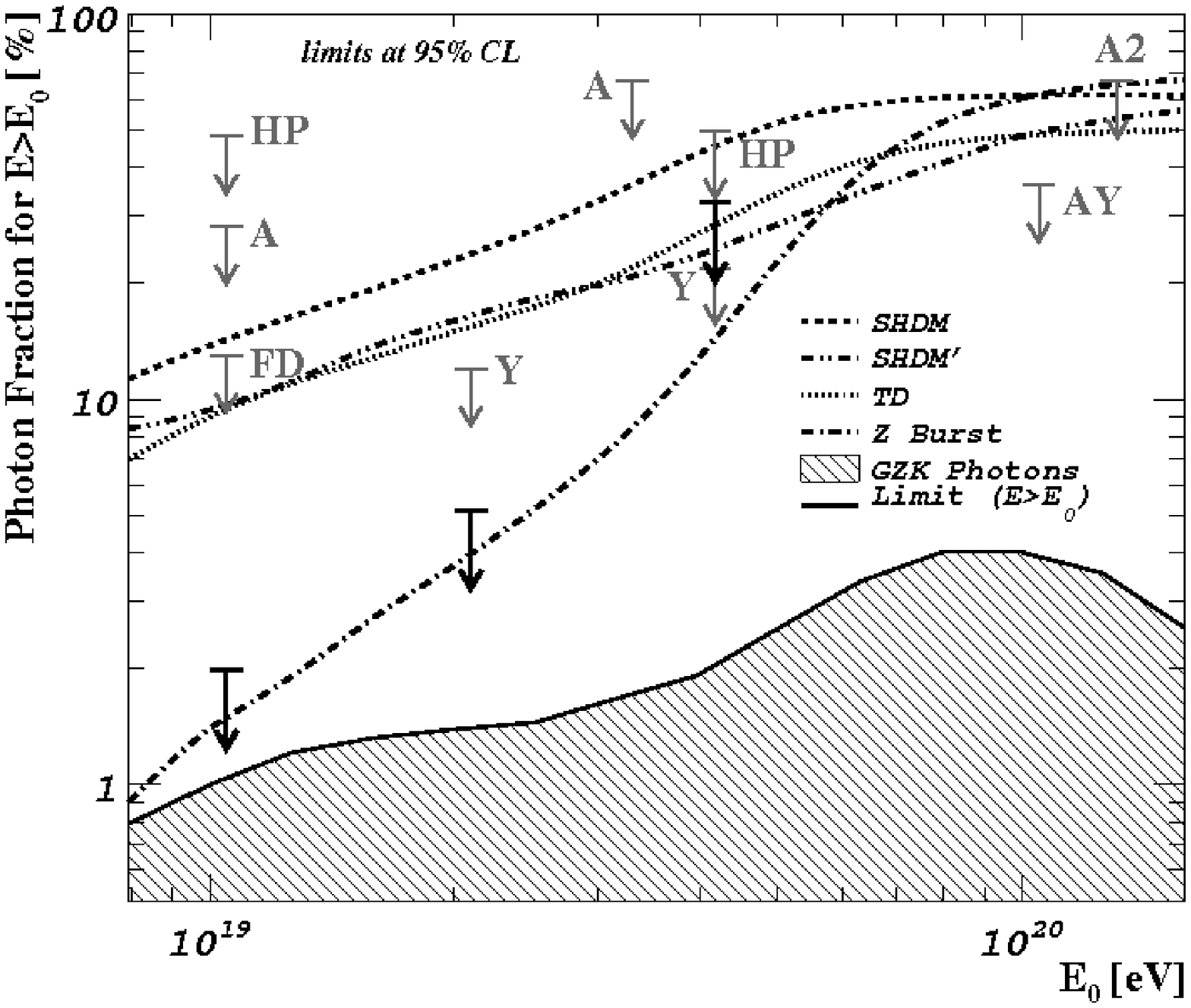}
\caption{\label{photonb}Upper limits on the fraction of photons in the
  integral CR flux compared to bounds from previous experiments. Lines indicate
  predictions in different top-down models. Shaded region is for GZK photons.}
\end{minipage}\hspace{1pc}%
\begin{minipage}{19pc}
\includegraphics[width=18pc]{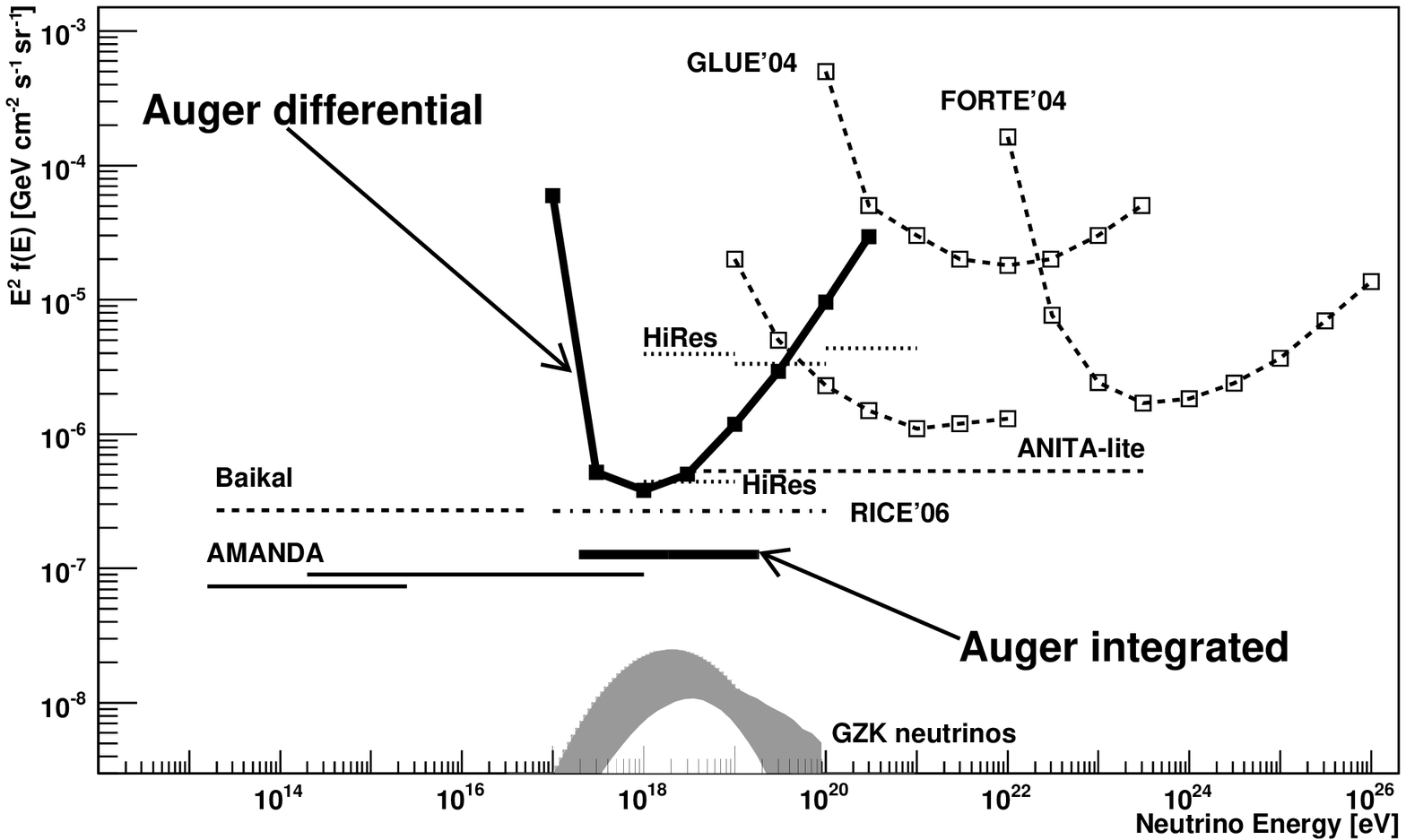}
\caption{\label{nub} 90\% CL bounds on diffuse flux of $\nu_\tau$ from Auger. 
Bounds from other experiments
apply assuming equal proportions of the three $\nu$ flavors  (horizontal lines
assume $E^{-2}$ spectrum).  Shaded region
  indicates possible expectations from GZK neutrinos.}
\end{minipage} 
\end{figure}

Also neutrinos have been searched with Auger trying to identify horizontal
showers with a significant electromagnetic component (i.e. young showers
produced by deeply penetrating particles). The background from
horizontal showers of hadronic origin would  have their electromagnetic
component attenuated well before reaching ground, being then dominated by
the muonic component that just produces narrow pulses in the SD
detectors. Searching then for elongated footprints on the ground
consistent with the horizontal propagation of the shower front 
at close to the speed of light and
 requiring that a large fraction of the 
triggered detectors have broad `electromagnetic rich' pulses 
allows then to identify the neutrino induced showers. No candidates of this
kind were found up to now. Although electron or tau neutrinos from slightly
above the horizon may produce this type of signals, the most sensitive search
is that for tau neutrinos from slightly below the horizon, since they can
interact in the rock and the tau leptons so produced can travel several tens of
km before decaying, and hence can be 
efficiently observed when the decays happen just above the detector.
Note that even if the sources
produce only muon and electron neutrinos by charged pion decays, due to
neutrino oscillations an equal
admixture of electron, muon and tau neutrinos
 would be expected upon arrival on Earth.
The bounds obtained for the tau neutrinos
should hence apply to all neutrino flavors. The present bounds are
shown in fig.~6 \cite{nubound} 
and the best sensitivity to $E^2 f(E)$ is around
$10^{18}$~eV, just where the GZK neutrinos produced in the photopion
interactions of the  CR protons are expected (grey region in the
plot). The sensitivity of the Auger instrument
 will steadily approach the relevant level of
fluxes (although if CRs are mostly heavy, that prediction moves down
significantly).

Finally there have also been some relevant anisotropy
studies at lower energies, around
the EeV. Being in the southern hemisphere, the Auger Observatory has a
privileged view towards the galactic centre (GC), 
which passes at just $6^\circ$
from the zenith at the site. 
This allowed to test claims from previous work 
 indicating possible excess fluxes from directions near it.
In particular, the  AGASA collaboration found
 a 4.5$\sigma$
excess ($observed/expected=506/413.6$)  in a $20^\circ$ radius region
for the energy range
$10^{18}$--$10^{18.4}$~eV, while for the same region and
energies Auger data led to $obs/exp=2116/2159.6$ \cite{gcenter}, a result 
 inconsistent with a large
excess. Similarly, an excess reported by the SUGAR collaboration in a $5^\circ$
region slightly displaced from the GC  was not confirmed
by Auger. A map of
overdensity significances  on $5^\circ$ radius windows in the region around
the GC is shown in fig.~7, together with the regions were the
AGASA and SUGAR excesses were reported. The excesses present
in this map 
are consistent with the expectations from fluctuations of an isotropic
distribution. 

\begin{figure}[h]
\begin{minipage}{17pc}
\includegraphics[width=17pc]{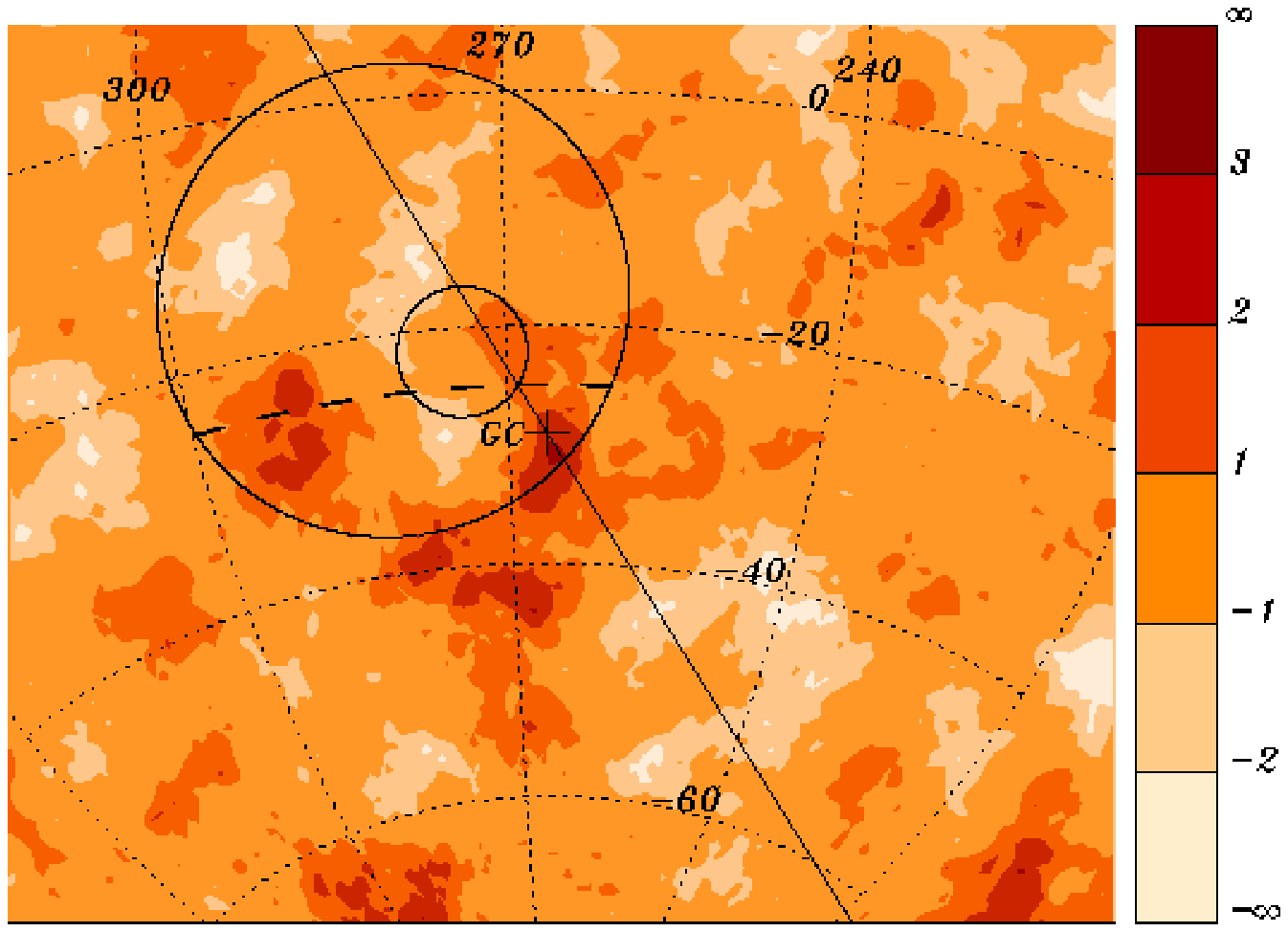}
\caption{\label{gcmap} Map in equatorial coordinates  around
  the GC (cross) showing the significance of the
  overdensities in $5^\circ$ radius windows, for $10^{17.9}\
  {\rm eV}<E<10^{18.5}$~eV. }
\end{minipage}\hspace{1pc}%
\begin{minipage}{20pc}
\includegraphics[width=18pc]{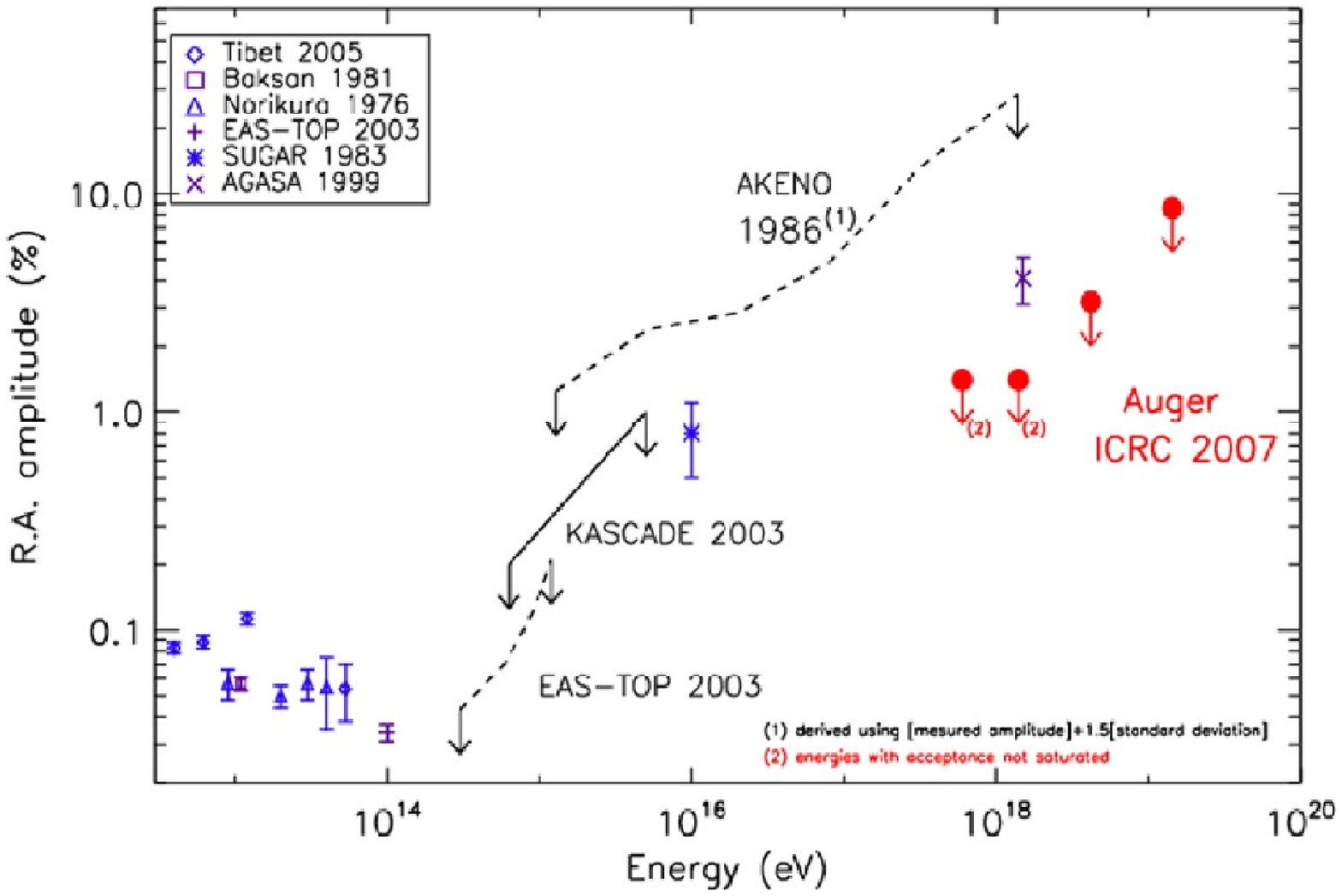}
\caption{\label{rabound} Summary of Auger 95\% CL upper bounds on the amplitude of a
  dipolar  modulation in right ascension and results from previous
  experiments. }
\end{minipage} 
\end{figure}

Another result from the AGASA collaboration
indicated the presence of a modulation in the right
ascension distribution of the events, with 4\% amplitude at EeV energies. The
search for a dipolar amplitude of this kind in data from the Auger Observatory
gave negative
results, allowing to set an upper bound on the amplitude of
 1.4\% at 95\% CL \cite{lscale},
 contradicting the previous finding. This kind of
 modulations could arise from the diffusion of galactic CRs out of the
 galaxy, and their search will allow to set constraints on the
 galactic/extragalactic transition. Moreover, two enhancements of the Auger
 Observatory are being done at present, one extending the field of view of the
 FD detectors to $60^\circ$ elevation above the horizon with new telescopes
 (HEAT project)  and the other being
 an infill of SD and muon detectors in a small part of the array  (AMIGA
 project). Both developments will allow
 the observation of showers down into the region of the
 second knee.

\end{document}